\documentstyle[prl,aps,multicol,epsf]{revtex}
\begin{document}
\widetext
\draft

\title{Double exchange-driven spin pairing at the (001) surface of manganites}
\author{Alessio Filippetti\thanks{Present address:
Materials Research Lab, Univ. of California, Santa Barbara, Ca 93106} 
and Warren E. Pickett}
\address{Department of Physics, University of California at Davis, 
Davis, California 95616}
\maketitle
\begin{abstract}
The (001) surface of La$_{1-x}$Ca$_x$MnO$_3$ system in various magnetic 
orderings is studied by first principle calculations. 
A general occurrence is that $z^2$ dangling bond charge -- which is
``invisible'' in the formal valence picture -- is promoted to the
bulk gap/Fermi level region.  This drives a double-exchange-like
process that serves to align the surface Mn spin with its subsurface
neighbor, regardless of the bulk magnetic order.  For heavy doping,
the locally ``ferromagnetic'' coupling is very strong and the moment
enhanced by as much as 30\% over the bulk value.
\end{abstract}


\begin{multicols}{2}

Although most efforts on the colossal magnetoresistance (CMR) materials
typified by the  La$_{1-x}$Ca$_x$MnO$_3$ (LCMO) system
are still concentrated on bulk 
properties, growing interest 
is being shown in the surface behavior\cite{park,peng,choi}.  
Knowledge of surface properties is essential not only to develop a 
perovskite manganite-based technology but also to determine fundamental
phenomena and mechanisms of magnetoelectronic behavior.
Indeed, the CMR effect occurs at high temperature,
around the magnetic ordering temperature, and a magnetic 
field of several Tesla 
is required to suppress the thermal magnetic disorder and produce the 
change in resistivity. Since high magnetic fields are generally 
unavailable in applications, alternative ways to trigger 
large low-field MR were
considered, such as with trilayer junctions\cite{tri} and 
polycrystalline samples\cite{poli}. The junctions are epitaxially grown 
along the [001] direction, and are made of a central insulating thin film of 
SrTiO$_3$ (the barrier), sandwiched by two metallic layers of 
La$_{0.67}$Sr$_{0.33}$MnO$_3$ (LSMO). Applying a low magnetic 
field, the tunneling conductivity can 
be switched by inducing a parallel (switch on) or anti-parallel 
(switch off) spin-orientation in the two electrodes.  Taking advantage 
of their half metallicity gives a very large tunneling MR (TMR).

Large low-field intergrain MR (IMR)\cite{IMR1} over a large temperature 
range has been
observed in polycrystalline samples of LSMO,\cite{IMR1,IMR2} 
CrO$_2$,\cite{IMR3} and the double 
perovskite systems Sr$_2$Fe(Mo,Re)O$_6$,\cite{IMR4,IMR5}
 all of which are expected to be half
metallic magnets. 
Magnetotunneling across grain boundaries, in which the relative orientation 
of the magnetization of neighboring grains is manuipulated by an
applied field, is believed to be the mechanism. 
In the IMR process, which may be the most promising for MR applications,
there is mounting evidence that the
state of the surfaces of the grains is important in the intergrain
tunneling process.\cite{IMR2,IMR3,IMR4}
For TMR it has long been clear that tunneling characteristics 
are strongly influenced, perhaps even dominated, by the electronic
and magnetic structure at the interface, and for IMR surface states
have been suggested to play the central role.

In the few experimental works present in the literature
intrinsic difficulties have been reported in the process of obtaining 
clean, bulk-truncated surfaces, due to surface segregation that occurs 
during growth at high temperature\cite{choi}, and strain effects induced 
by film-substrate mismatch\cite{peng}. Structural and electronic 
properties of the low-index surfaces (including the possibility 
of reconstructions) are still unknown, in spite of their importance in
establishing the half metallic nature of the CMR materials using
photoelectron emission.\cite{park} 
However, advancements in epitaxial growth and surface uniformity are
being reported,\cite{kawasaki} so a first fundamental step towards 
describing real surfaces consists in understanding how the 
intrinsic properties of the ideal unreconstructed surfaces differ from the 
respective bulk properties, i.e. how the bulk truncation in itself 
modifies the physics of this compounds.
First-principle calculations are ideally suited for this aim, and
in this paper we focus on surface spin ordering of the Mn-terminated (001) 
surface of La$_{1-x}$Ca$_x$MnO$_3$. 
Although this system shows an extremely rich variety of magnetic phases 
for different level of doping\cite{book}, 
we identify a general, robust mechanism that should dominate the surface
spin order for any doping level $x$.  In the regime of heavy
doping $x\sim  1/2$, the surface-to-subsurface magnetic 
coupling is much stronger
than in bulk, and the surface moment is enhanced by as much as 30\%.

Based on the growing understanding of the double exchange (DEX)
process in bulk manganites,\cite{book}
it can be expected that the surface spin alignment will be strongly
dependent on the Mn $e_g$ occupation.  At the (001) surface, however, the
$e_g$ degeneracy is broken: the $x^2-y^2$ orbital remains
very strongly $dp\sigma$ hybridized with neighboring (in surface layer)
O ions, but the $z^2$ orbital is left ``dangling.''
The implications of this symmetry breaking were first glimpsed in the
simple, undoped $x$=1 member CaMnO$_3$, which has G-type AFM bulk ordering 
due to standard AFM superexchange between filled t$_{2g}$ shells.
These $t_{2g}$ shells contain the nominal three electrons assigned to
Mn$^{4+}$ in the formal valence picture, with the $e_g$ formally empty.
As it has been pointed out in other contexts,\cite{roncohen}, 
the amount of actual
$d$ charge is not at all identical to the formal $d^n$ charge.  For
bulk phenomena however, this idealization usually gives a reliable
broad picture of general behavior, including spin, charge, and orbital
order.  For the CaMnO$_3$ (001) surface, however, G-type spin order does not
survive at the surface.  Instead, a flip of all spins in the surface layer 
occurs, driven by the appearance of Mn $z^2$ charge\cite{fp} that drives a
double exchange (DEX) process that strives to align spins.  This $z^2$
charge is present in the bulk, resulting from $dp\sigma$ mixing that
draws a truly significant amount of $e_g$ charge into the ``O 2$p$"
bands: the 18 O $p$ bands actually contain on the order of 1.5-2 electrons
of Mn $e_g$ character.  Some of this becomes a dangling bond band at
the surface, lying in the bulk gap and driving the surface spin to flip.
What we show in this paper is that this mechanism survives, and in fact
is enhanced, as doping occurs. 

\begin{figure}
\epsfxsize=5cm
\centerline{\epsffile{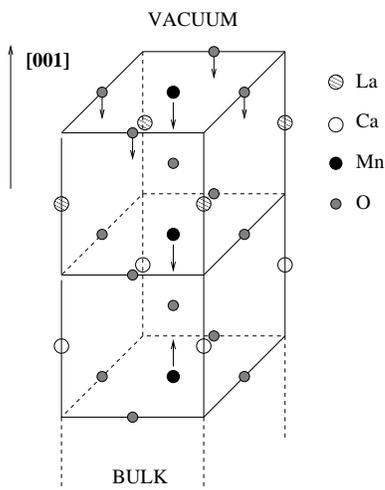}}
\caption{Structure of the (001) surface of
tetragonal La$_{1/2}$Ca$_{1/2}$MnO$_3$. Arrows indicate spin
orientations for the stable $\uparrow \downarrow \downarrow$ ordering
(see text), which we find for this compound and for
CaMnO$_3$.  Parallel alignment of the surface and subsurface layers is
expected to be true generally.  Note that the surface oxygen ions are
also polarized.}
\label{fig1}
\end{figure}

First-principles calculations have been performed within local-density
approximation (LDA), employing a plane wave basis and Vanderbilt 
pseudopotentials\cite{van}. A 30 Ryd cut-off energy and the 
exchange-correlation potential of Perdew and Zunger\cite{PZ} was used.
For the La$_{1-x}$Ca$_x$MnO$_3$ (001) surfaces we have studied, we used 
slab of nine atomic layers. 
For $x$=1/2, the stacking along $\hat{z}$ is made by 
alternating layers of La and 
Ca (see Fig. \ref{fig1}), retaining a mirror symmetry with respect 
to the central 
Mn layer. 
The artificial ordering of La and Ca layers (which must be done somehow
in a finite
supercell) should not
affect our conclusions, since they simply become ionized by contributing
their valence electrons to the O and Mn bands.
For the planar lattice constant of bulk La$_{1/2}$Ca$_{1/2}$MnO$_3$
we obtain by energy minimization a$_0$= 7.21 a.u, which is a reasonable 
value between the experimental 7.35 a.u. for La$_{2/3}$Ca$_{1/3}$MnO$_3$, 
and 7.05 a.u. for CaMnO$_3$. We find the AFM phase favored by 
15 meV/Mn over the FM, which has a nearly half-metallic density 
of states. 

In Table \ref{tavola} the calculated relative surface energies for $x$=1/2 are 
reported. As can be seen from Fig. 1, there are two kinds 
of Mn-terminated surfaces, i.e. one with 
La in the subsurface layer (indicated as Mn-La), and one with Ca instead
(Mn-Ca). Since (Table I) they give equivalent results, we will quote
specific results for just one of them (La-Mn).  Treating an inner region
with the true spin, charge, and orbital order for $x$=1/2\cite{st}
is well beyond
computational capabilities, but the behavior we identify is so robust
that we expect it to be independent of bulk order.  For 
each of the geometries of our nine layer surface, 
four spin arrangements on Mn are possible, 
labeled in Table \ref{tavola}
by arrows on central (C), subsurface (SS) and surface 
(S) Mn, in this order. These are: surface 
and subsurface parallel, aligned or antialigned to the central layer,
and surface and subsurface antiparallel, with the subsurface aligned or
antialigned with the central layer.
We find that S-SS spin alignment is strongly favored, with
the most stable configuration having both the surface and subsurface
layer spins antiparallel to the central Mn spin:
(from third to top layer) $\uparrow\downarrow\downarrow$. 

The energies can be mapped onto an interlayer Ising model with 
three independent 
effective exchange constants: J$_{S-SS}$, J$_{SS-C}$ and J$_{S-C}$, the
latter being a second neighbor coupling.  J$_{SS-C}$ = -18 meV 
(AFM) is close to the exchange parameter obtained directly 
from the bulk calculation (J$_{bulk}$ = -15 meV).  
The interaction between Mn on first and third layers, 
J$_{C-S}$ = 8 meV, is
FM in sign and is
related to the d$_{z^2}$ the surface state discussed below.

The most striking result for $x$=1/2 (Table \ref{tavola}) is the positive,
unusually large value of J$_{S-SS}$ = 53 meV, more than three times larger
than, and opposite in sign to, the bulk AFM coupling.
For comparison, for CaMnO$_3$ ($x$ = 1) the interlayer exchange constant 
at the surface was 
29 meV.  (The bulk coupling for $x$=1 is also different from $x$=1/2, 
with J$_{bulk}$ = -26 meV.)
This large FM coupling, for both $x$=1 and $x$=1/2, is the
consequence of a very general characteristic of (001) surface formation.

\begin{figure}
\epsfxsize=6cm
\centerline{\epsffile{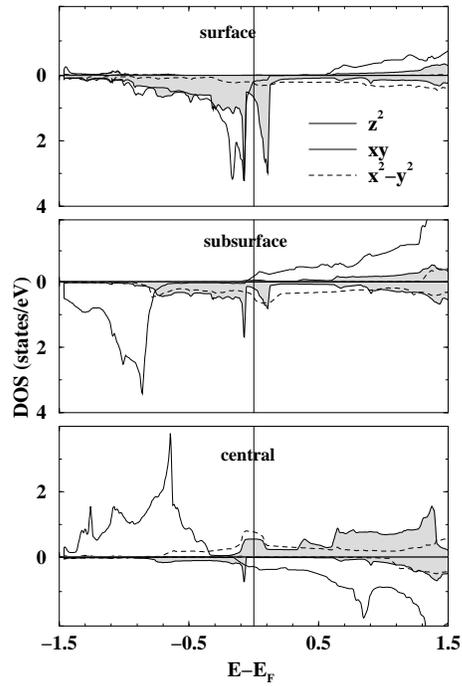}}
\caption{Orbital-resolved  Mn $d$ DOS for the (001) 
Mn-La surface in the spin configuration $\uparrow\downarrow\downarrow$ 
(i.e. AFM bulk and flipped spin at surface). The panels 
refer to the 3 unequivalent Mn layers in the slab.
\label{fig2}}
\end{figure}

In Fig.\ref{fig2} the orbital-resolved density of states (DOS) of 
the Mn ions
for the (001) surface in the most stable spin configuration 
(i.e. $\uparrow\downarrow\downarrow$) is shown. 
Two surface Mn d$_{z^2}$ DOS peaks 
straddle the Fermi energy (E$_F$ = 0), with
a tail of occupied states that extends down to $\sim$ -1.5 eV.
These states are also visible on subsurface Mn and, for the
occupied peak, on central Mn as well.  Thus the surface has a
deep surface d$_{z^2}$ resonance extending to the fifth layer 
below the surface. 
In the majority channel of the central (`bulk')
Mn ion, d$_{z^2}$ and d$_{x^2-y^2}$ orbitals 
contribute to the DOS at E$_F$, whereas in the minority channel the only 
contribution comes from $t_{2g}$ states. 
The d$_{z^2}$ dangling bond discussed above 
leads to the formation of the 
surface resonance as it does in CaMnO$_3$. 
It is also apparent that the d$_{xy}$ bands are shifted upward 
in energy, so that the minority channel is depleted (i.e. the 
Mn at surface is fully polarized) and the d$_{xy}^{\uparrow}$ surface bands 
contribute to the DOS at E$_F$. The magnetic moment on the surface Mn 
(3.23 $\mu_B$) is 10\% larger than on subsurface Mn (2.97 $\mu_B$) 
and 30\% larger than in `bulk' central Mn (2.50 $\mu_B$), 
but the total charge on Mn 
($\sim$ 5.3 electrons using our definition) 
is nearly the same at surface and in the bulk.
The increase of magnetization is due mostly to the 
d$_{z^2}$ polarization with some contribution from 
the depletion of d$_{xy}^{\uparrow}$ states around  E$_F$.
Also, a small intra-atomic charge readjustment occurs from 
d$_{x^2-y^2}$ and d$_{xy}$ to the polarized d$_{z^2}$ orbital on
the surface Mn ion.

\begin{figure}
\epsfxsize=4cm
\centerline{\epsffile{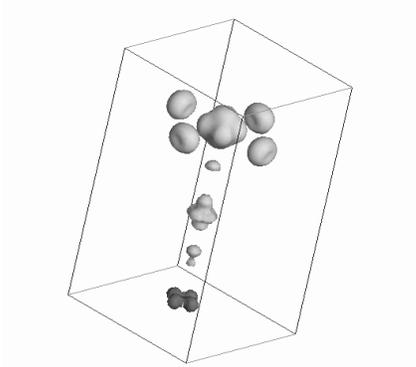}}
\caption{Isosurfaces of the valence magnetization for the (001) Mn-La 
surface in the spin configuration $\uparrow\downarrow\downarrow$.
The magnetization shown is due to states below, and within, 
0.3 eV of E$_F$ (see the corresponding DOS in Fig.\ref{fig2}). 
Dark and light isosurfaces are of same magnitude but opposite sign, 
i.e. they represent up and down spin densities, respectively.
\label{fig3}}
\end{figure}

The resulting surface polarization can be visualized from the
isosurfaces of the magnetization displayed in Fig.\ref{fig3}. 
Contributions coming from states that lie in the region 
within 0.3 eV below E$_F$ are shown, {\it i.e.} the ``core'' $t_{2g}$ 
moments are not included in the subsurface and central Mn, whereas 
the surface Mn shows a combination of d$_{z^2}$ and d$_{xy}$ spins; 
on subsurface Mn the d$_{z^2}$ magnetization is mixed with some 
d$_{x^2-y^2}$. The double exchange effect between d$_{z^2}$ orbitals on 
surface and subsurface Mn comes into play and
leads to the strong FM coupling J$_{S-SS}$=53 meV responsible for the 
spin alignment. On the central Mn with its antialigned spin,
the magnetization is
d$_{x^2-y^2}$-like.  (Unfortunately, present computational limitations
do not allow us to study an eleven layer slab, for which the central
layer should be more bulklike.)
Also evident in Fig. \ref{fig3} is that a remarkably large fraction 
of this surface-induced magnetization lies in the 
O $p_{\pi}$ orbitals of the surface layer.  Polarization of the O ion in FM 
bulk environments in manganites has been emphasized elsewhere.\cite{wepdjs}

The change of the Mn d$_{z^2}$ orbital from broad, 
strongly $dp\sigma$ hybridized
in the bulk to an atomic-like, narrow in energy, surface state
is a very specific feature 
of this (001) surface formation, and this surface dehybridization
generally should be described well by LDA. 
We suggest that this effect is strong enough to turn the AFM spin 
coupling into FM for any doping level.
At least two arguments support this hypothesis.  First, the spin-pairing
occurs for the (001) surface of CaMnO$_3$\cite{fp} that should be the
most unfavorable case, since in the bulk (nominally) only the 
majority t$_{2g}$ orbitals are occupied, thus their AFM character is 
dominant. Nevertheless, the partially occupied d$_{z^2}$ surface state 
reverses the magnetic coupling. Second, the very large
change of exchange interaction parameter (from -15 meV in bulk
to +53 meV at the surface)  
would overcome AFM bulk coupling even stronger than the one
considered here. 

A crucial case is the $x$=0 member LaMnO$_3$, which is A-type AFM in the bulk.
The spin-pairing argument applied to the surface parallel to the FM (001) 
layers predicts a spin-flip of the surface Mn layer. The AFM spin coupling along
the $\hat{z}$ axis is robust and explained by a well estabilished picture:
the in-plane FM coupling is stabilized by the 
ordering of Mn $e_g$ orbitals, so that occupied d$_{x^2}$ (d$_{y^2}$) 
orbitals alternates with empty d$_{x^2}$ (d$_{y^2}$) orbitals on 
neighboring Mn. Thus, all the $e_g$-type charge fills in-plane orbitals,
and the d$_{z^2}$ orbitals are empty and higher in energy. As a consequence,
the AFM interactions between neighboring t$_{2g}$'s dominates in the 
orthogonal direction.
A  realistic first-principle calculation of the LaMnO$_3$ surface is beyond
the possibility of detailed calculations, since it would require a 
$\sqrt 2 \times \sqrt 2$ lateral enlargement of the cell as well as additional 
thickness to treat the tilting of the MnO$_6$ octahedra\cite{elemans}, and
the Jahn-Teller distortion at the surface would have to be determined. 
However, the formation of the d$_{z^2}$ surface state within bulk LaMnO$_3$ 
gap seems to be beyond doubt, based on the behavior of the d$_{z^2}$ dangling
bond for $x$=1 and $x$=1/2. The question is whether this would be able to
overcome the t$_{2g}$ AFM contribution. 
In Ref.\onlinecite{sht} it is shown that 
the t$_{2g}$ contribution for bulk LaMnO$_3$ increases linearly in magnitude 
with the distortion between in-plane and inter-planar lattice constants
at fixed volume, i.e. the AFM coupling between (001) planes increases linearly
by shortening the interplanar distance, likely due to the electrostatic
repulsion that further depletes the d$_{z^2}$ orbitals.  
The t$_{2g}$ contribution to J$_{bulk}$ has been calculated in 
Ref.\onlinecite{sht} 
as a function of the lattice distortion. It is in the range of $\sim$ 20-30 meV, 
i.e. not large enough to overcome the value of J$_{S-SS}$, thus we definitely
expect the occurrence of a spin-flip process at the (001) LaMnO$_3$ surface.

\begin{figure}
\epsfxsize=4cm
\centerline{\epsffile{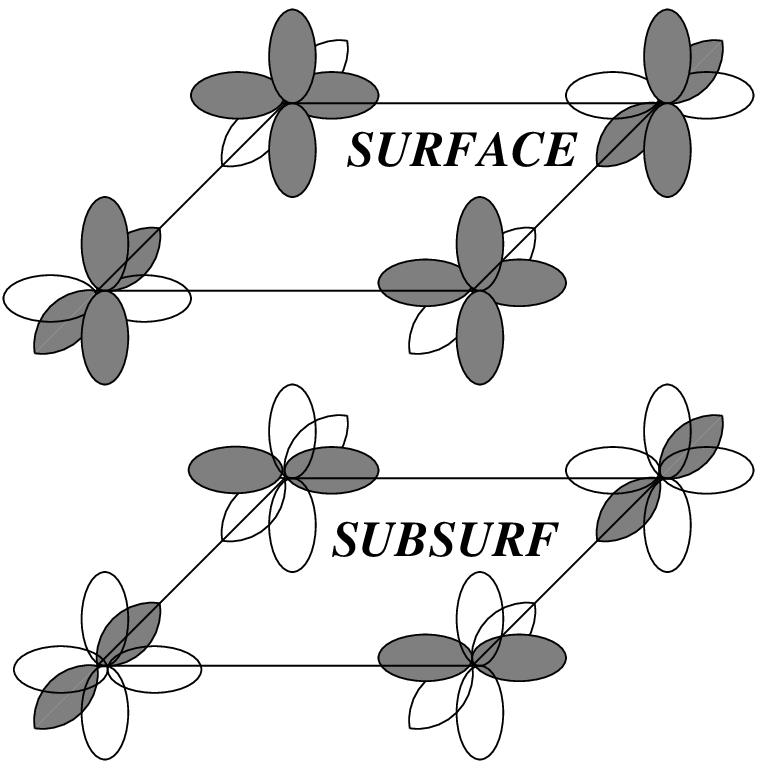}}
\caption{Orbital ordering at LaMnO$_3$ (001) surface: filling 
of d$_{z^2}$ orbital (indicated by shading) at the surface produces 
FM spin coupling perpendicular
to the surface.  Planar orbital ordering in the subsurface and other
buried layers leads to FM layers alternating in spin direction, except
at the surface.
\label{fig4}}
\end{figure}

In Fig.\ref{fig4} we show the $e_g$ orbitals on surface and subsurface
layers, and indicate the expected filling and orbital ordering after
the formation of the surface state.
The orbitals are ordered in ``FM'' fashion both in-plane and
orthogonally to the surface, as a result of the surface formation that fills 
the Mn surface d$_{z^2}$ orbital no longer degenerate with the d$_{z^2}$ 
orbital of the underliyng subsurface Mn.

To summarize, we have found that terminating the (001) surface
of La$_{1-x}$Ca$_x$MnO$_3$ surfaces with the Mn ion exposed, results
in partial filling of the d$_{z^2}$ orbital that drives a
double-exchange-like ordering of the surface and subsurface layers
of Mn ions. We have shown this effect explicitly for $x$=1/2 and
(previously) for undoped CaMnO$_3$. A comparison between these two cases
indicates that it is stronger in doped systems.  
This result has important implications (1) for surface studies, where 
this effect tends to insure that surfaces of the CMR materials 
($x\approx$1/3) will remain ferromagnetically aligned and half metallic 
as well, as supported by photoemission studies, and (2) for the intergrain
magnetoresistance effect, where the magnetic structure of the grain
surfaces can strongly affect the device characteristics.
This behavior, which is strongly related to band filling but much less
dependent on ion size effects, should also hold for the
La$_{1-x}$Sr$_x$MnO$_3$ and La$_{1-x}$Ba$_x$MnO$_3$ systems.

This research was supported by National Science Foundation grant
DMR-9802076.  Calculations were done at the Maui High Performance
Computing Center.

\narrowtext

\begin{table}
\centering\begin{tabular}{cccccccc}
  C-SS-S & $\uparrow\uparrow\uparrow$ & $\uparrow\uparrow\downarrow$ &
  $\uparrow\downarrow\uparrow$ &   $\uparrow\downarrow\downarrow$ &
  J$_{S-SS}$ & J$_{SS-C}$ & J$_{S-C}$ \\
\hline
  Mn-La        & 23 & 144 & 91 & 0 & 53 & --18 & 8 \\
  Mn-Ca        & 17 & 142 & 88 & 0 & 53 & --18 & 9 \\     
\end{tabular}
\caption{Energies (in meV) for different spin configurations on Mn atoms.
Each of them is labeled by the three arrows indicating the spin direction 
of central-subsurface-surface Mn. Mn-La is the Mn-terminated (001) surface 
with La on second layer, Mn-Ca is that one with Ca on second layer.
All energies refers to that one of the most stable arrangement, i.e.
$\uparrow\downarrow\downarrow$. 
\label{tavola}}
\end{table}

\end{multicols}





\begin{thebibliography}{99}

\bibitem{park}
J.-H. Park {\it et al.}, Phys. Rev. Lett. {\bf 81}, 1953 (1998).

\bibitem{peng}
H. B. Peng {\it et al}., Phys. Rev. Lett. {\bf 82}, 362 (1999).

\bibitem{choi}
J. Choi {\it et al}, Phys. Rev. B {\bf 59}, 13453 (1999).

\bibitem{tri}
Y. Lu. {\it et al.},  Phys. Rev. B {\bf 54}, 8357 (1996);
J. Z. Sun {\it et al.}, Appl. Phys. Lett. {\bf 70}, 1769 (1997).

\bibitem{poli}
A. Gupta {\it et al.}, Phys. Rev. B {\bf 54}, 15629 (1996).

\bibitem{IMR1}H. Y. Hwang {\it et al.}, Phys. Rev. Lett. {\bf 77}, 2041 (1996).

\bibitem{IMR2}S. Lee {\it et al.}, Phys. Rev. Lett. {\bf 82}, 4508 (1999).

\bibitem{IMR3}H. Y. Hwang and S.-W. Cheong, Science {\bf 278}, 1607 (1997).

\bibitem{IMR4}K.-I. Kobayashi {\it et al.}, Phys. Rev. B {\bf 59},
 11159 (1999).

\bibitem{IMR5}T. H. Kim {\it et al.}, Appl. Phys. Lett. {\bf 74}, 1737 (1999).

\bibitem{kawasaki}M. Kawasaki {\it et al.}, Mat. Sci. and Eng. B {\bf 63},
49 (1999).

\bibitem{book}
{\it Physics of Manganites}, edited by T. A. Kaplan 
and S. D. Mahanti (Kluwer/Plenum, New York, 1999).

\bibitem{roncohen}R. E. Cohen, Nature {\bf 358}, 136 (1992).

\bibitem{fp}
A. Filippetti and W. E. Pickett, Phys. Rev. Lett. {\bf 83}, 4184 (1999).

\bibitem{van}
D. Vanderbilt, Phys. Rev. B {\bf 32}, 8412 (1985);
K. Laasonen {\it et al.}, Phys. Rev. B {\bf 47}, 10142 (1993).

\bibitem{PZ}J. P. Perdew and A. Zunger, Phys. Rev. B {\bf 23}, 5048 (1981).

\bibitem{st}
I. V. Solovyev and K. Terakura, Phys. Rev. Lett. {\bf 83}, 2825 (1999).

\bibitem{wepdjs} W. E. Pickett and D. J. Singh, Phys. Rev. B {\bf 53},
1146 (1996).

\bibitem{sht}
I. Solovyev, N. Hamada, and K. Terakura, Phys. Rev. Lett. 
{\bf 76}, 4825 (1996).

\bibitem{elemans}J. B. A. A. Elemans {\it et al.}, J. Solid State Chem.
{\bf 3}, 238 (1971).

\end{thebibliography}
\end{document}